\documentclass[aps,prb,twocolumn,longbibliography,superscriptaddress,floatfix,noeprint,raggedbottom]{revtex4-2}
\usepackage{bm}
\usepackage{color} 
\usepackage{graphicx}
\usepackage{epsfig}
\usepackage{hyperref}
\usepackage{natbib}
\usepackage{amsmath}
\usepackage{amssymb}
\usepackage{bbm}
      
\newcommand {\e}{{\rm e}}

\begin{document}

\title{Selective spin-caging in diffusive networks}

\author{A.~V.~Poshakinskiy}
\email[E-mail:~]{poshakinskiy@gmail.com}
\affiliation{Ioffe Institute, St. Petersburg 194021, Russia}

\begin{abstract}

We propose a semiclassical analogue of Aharonov--Bohm caging based on spin diffusion in spin-orbit-coupled semiconductor networks. We present a feasible rhombic design for one- and two-dimensional periodic networks where Rashba and Dresselhaus spin-orbit interactions cause spins propagating along two alternative paths to arrive with opposite orientations, completely suppressing their diffusion and drift. Crucially, this cancellation occurs only for certain spin orientations, while the others propagate freely, allowing the network to act as a perfect spin filter.

\end{abstract}

\maketitle

\section{Introduction}

Aharonov--Bohm caging is an extreme form of localization in which destructive interference confines a wave packet to a finite cluster of an otherwise extended periodic network~\cite{Vidal1998}. In suitable lattices, a magnetic flux of half a flux quantum per plaquette produces a relative phase of $\pi$ for all pairs of alternative paths. Their amplitudes then cancel, blocking transport and leading to flat energy bands. In real structures, complete caging is hindered by electron scattering that spoils coherence, though small signatures still remain in magnetoresistance~\cite{Vidal2000,Naud2001}.

Instead of a magnetic field, other kinds of gauge fields can be used. Caging effects have been realized in photonic waveguide arrays~\cite{Mukherjee2018Cages} and momentum-space lattices of ultracold atoms~\cite{Li2022Cages} with synthetic gauge fields, and proposed in spin-orbit-coupled exciton-polariton lattices~\cite{Qi2025Cages}. For electrons in semiconductors, $k$-linear spin-orbit interaction can be regarded as an $\mathrm{SU}(2)$ gauge field acting on the electron spinor~\cite{Tokatly2010a}. Interference of spinor amplitudes in coherent quantum-wire networks~\cite{Aronov1993,Nitta1999,Koga2004,Konig2006,Frustaglia2004,Hijano2021} was proposed to achieve caging effect~\cite{Bercioux2004,Mukherjee2019SpinSelective}. However, the interference still requires ballistic electron propagation and is destroyed by any scattering event. For diffusive electron propagation, the spin-orbit caging effect is reduced to a small weak-(anti)localization-like correction to the Drude conductivity~\cite{Mathur1992,Bergsten2006,Zhu2007,Nagasawa2012,Nagasawa2013,Nagasawa2018,Wang2019SpinInterferometry,Poshakinskiy2026arxiv}.

\begin{figure}[b!]
    \centering
    \includegraphics[width=0.99\columnwidth]{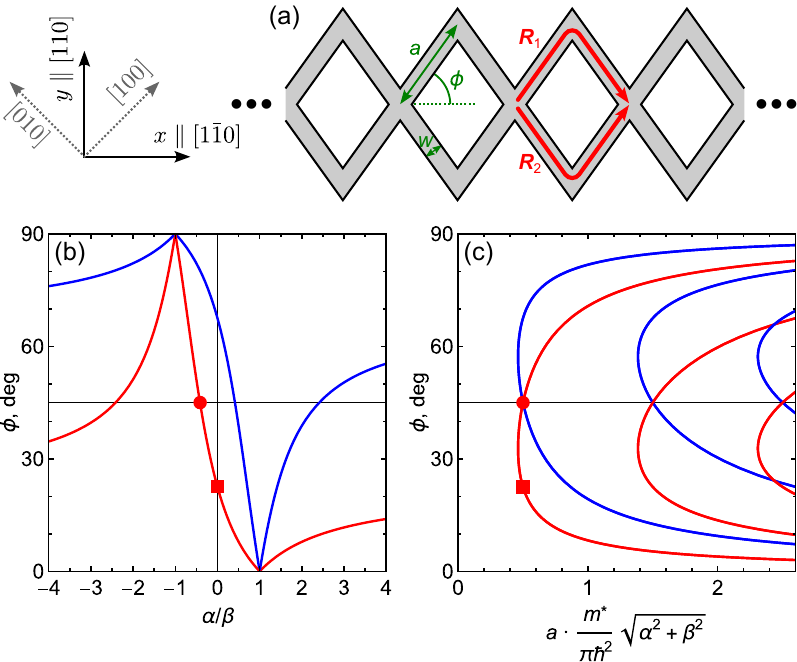}
    \caption{(a) Schematic of a diamond-shaped spin-diffusion network. The two possible paths connecting a pair of adjacent vertices, and the corresponding spin rotation matrices $\bm R_1$ and $\bm R_2$, are shown in red. (b,c) Relation between the spin-orbit coupling parameters $\alpha$, $\beta$ and the network geometry parameters $\phi$, $a$ that enables the spin-caging effect. The red circle and square indicate the parameters chosen for calculations in this paper.
}
    \label{fig:pars}
\end{figure}

Here we propose a semiclassical analogue of Aharonov--Bohm caging for electron spin polarization in periodic networks of diffusive semiconductor channels. Inside the channels, the Dyakonov-Perel spin relaxation is suppressed~\cite{Malshukov2000,Kiselev2000,Kunihashi2012} and the spin-orbit coupling rotates the spin as an electron moves along a channel~\cite{Altmann2014,Altmann2015,Ishihara2014Wires,Grobecker2025}. The rotations accumulated along different links of the network generally do not commute, and different paths on the network can lead to different spin rotations~\cite{Poshakinskiy2025}, enabling one to engineer a caging effect. In contrast to coherent electron transport, spin polarization in diffusive transport is a three-component vector, and the effect of spin-orbit interaction is described by an $\mathrm{SO}(3)$ gauge field. Caging requires the spin polarization vectors to cancel after diffusion along two alternative paths through the network, described by the $\mathrm{SO}(3)$ matrices $\bm R_1$ and $\bm R_2$, see Fig.~\ref{fig:pars}(a),
\begin{align}\label{eq:cagecond}
(\bm R_1 +\bm R_2)\bm S_0 = 0 \,.
\end{align}
However, this is impossible for all initial spin orientations $\bm S_0$ because it would require $\bm R_1^{-1}\bm R_2=-\bm 1$, which is not an $\mathrm{SO}(3)$ rotation. Instead, the caging condition Eq.~\eqref{eq:cagecond} can hold for two spin directions, say $x$ and $y$, while being violated for the remaining $z$ component. We show that such selective caging can be achieved in (001)-oriented quantum wells with Rashba and Dresselhaus spin-orbit coupling and that it enables efficient spin filtering.

\section{Model}

We consider a degenerate two-dimensional electron gas with spin-orbit interaction $H_{\rm SOI}=\bm h(\bm k)\cdot\bm\sigma$, where $\bm\sigma$ is the vector of Pauli matrices and $\bm k$ is the electron wave vector. For a (001)-oriented zinc-blende quantum well, with $\bm e_x\parallel[1\bar{1}0]$, $\bm e_y\parallel[110]$, and $\bm e_z\parallel[001]$, the spin-orbit field on the Fermi circle consists of the first and third angular harmonics, $\bm h=\bm h_1+\bm h_3$~\cite{Ganichev2014,Salis2014}, with
\begin{align}\label{eq:SOI}
\bm h_1(\bm k)&= (\beta+\alpha)k_y\bm e_x+(\beta-\alpha)k_x\bm e_y\,, \\
\bm h_3(\bm k)&=\frac{\beta_3}{k^2}
\bigl[k_y(3k_x^2-k_y^2)\bm e_x +k_x(3k_y^2-k_x^2)\bm e_y\bigr] \,,
\end{align}
where $\beta=\beta_1-\beta_3$, $\alpha$ is the Rashba parameter, and $\beta_1$ and $\beta_3$ are the linear and cubic Dresselhaus parameters, respectively.
 
For diffusing electrons, the spin-orbit field rotates the spin as the electrons move.
The evolution of the spin density $\bm S(\bm r,t)$, $\bm r=(x,y)$, is described by the spin diffusion equation~\cite{Kalevich1994,Malshukov2000,Froltsov2001,Gridnev2002,Kleinert2007,Poshakinskiy2016cor,Ferreira2017,Schliemann2017,Passmann2019}
\begin{align}\label{eq:diff}
\frac{\partial \bm S(\bm r,t)}{\partial t }= \left[ D_s \sum_{\nu=x,y}\left(  \frac{\partial}{\partial r_\nu} - \bm \Lambda_{\nu} \right)^2  - \bm \Gamma' \right] \bm S(\bm r,t) \,,
\end{align}
where $D_s$ is the spin diffusion coefficient and the skew-symmetric 3$\times$3 matrices $\bm\Lambda_\nu$ describe the rotation of moving spins due to $\bm h_1(\bm k)$ and are defined by $\bm\Lambda_{x(y)}\bm S=(2m^*/\hbar^2)(\beta\mp\alpha)\bm e_{y(x)}\times\bm S$.
The combination $\partial/\partial r_\nu-\bm\Lambda_\nu$ in Eq.~\eqref{eq:diff} can be regarded as a covariant derivative, and the matrices $\bm\Lambda_\nu$ thus form an $\mathrm{SO}(3)$ connection. Since $\bm\Lambda_x$ and $\bm\Lambda_y$ do not commute in general, their combined action leads to Dyakonov--Perel spin relaxation~\cite{Tokatly2010a}.

An additional, typically small, spin-relaxation term $\bm\Gamma'=\operatorname{diag}(\Gamma',\Gamma',2\Gamma')$ in Eq.~\eqref{eq:diff} arises from the third angular harmonic of the spin-orbit field, with $\Gamma'=4D_s(m^*\beta_3/\hbar^2)^2$~\cite{Salis2014}. Random effective magnetic fields can also contribute to this relaxation~\cite{Sherman03}.

\begin{figure}[t!]
    \centering
  \includegraphics[width=0.93\columnwidth]{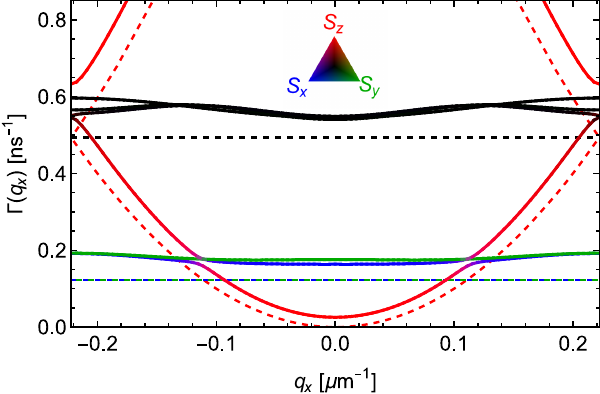}
    \caption{Dispersion of the Bloch-mode decay rates in the 1D rhombic structure shown in Fig.~\ref{fig:pars}(a), with the geometric and spin-orbit parameters corresponding to the red circle in Fig.~\ref{fig:pars}(b,c): $\phi=45^\circ$, $a=10\,\mu$m, $\alpha = 0.68\,$meV$\cdot$\AA, $\beta=-1.65\,$meV$\cdot$\AA. Other parameters are $D_s=200\,$cm$^2$/s, $w=2\,\mu$m, and $\beta_3=0.1\,$meV$\cdot$\AA.
    Color encodes the spin orientation at the network vertices. Solid lines show the numerical calculation; dashed lines correspond to an idealized network model that neglects finite channel width and cubic Dresselhaus spin-orbit interaction.}
    \label{fig:disp}
\end{figure}

\section{Spin caging condition}

\begin{figure*}[tb!]
    \centering
  \includegraphics[width=0.99\textwidth]{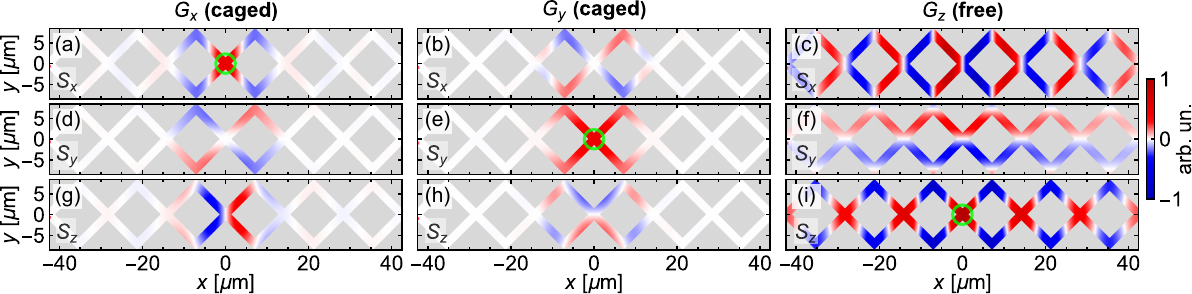}
    \caption{Stationary spin distributions in the 1D rhombic network under continuous local spin pumping at the center of the plot. Columns correspond to pumping of the different spin components $G_x$, $G_y$, and $G_z$. Rows show the spatial distributions of the spin-density components $S_x$, $S_y$, and $S_z$ for those cases.
 The pump profile was chosen to be Gaussian, with the half-maximum indicated by the green circles.
Network parameters are the same as in Fig.~\ref{fig:disp}.}
    \label{fig:dif1}
\end{figure*}

For the spin propagating from a vertex of the network, we want to cage the $x$ and $y$ spin components while leaving the $z$ spin component free. This requires that, when the spin propagates along the two possible paths between adjacent vertices, its $z$ component is preserved while the $x$ and $y$ components are eliminated, that is,
\begin{align}\label{eq:cond}
\tfrac12(\bm R_1+\bm R_2) = \bm P_z 
\end{align}
where the rotation matrices $\bm R_{1,2}$ are expressed as path-ordered exponentials along the two paths, $\bm R_{1,2} = \mathcal{P}\exp \int_{1,2}\sum_\nu \bm\Lambda_\nu dr_\nu$~\cite{Poshakinskiy2025}, and $\bm P_z = \bm e_z \otimes \bm e_z$ is the projector onto the $z$ axis. Condition Eq.~\eqref{eq:cond} can be satisfied only if both $\bm R_1$ and $\bm R_2$ are rotations about the $z$ axis, $\bm R_1 =\bm R_z^{\theta}$ and $\bm R_2 =\bm R_z^{\pi+\theta}$. This also ensures that the rotation around a plaquette is $\bm R_{\circlearrowright} = \bm R_2^{-1} \bm R_1 = \bm R_z^\pi$, which yields $-1$ for spin along $x$ or $y$, in agreement with the caging definition.

In addition to the above condition on the rotations, the diffusion probabilities for the two paths must be equal. The simplest way to ensure this is to choose a geometry in which the paths are related by a symmetry. In particular, if the two paths are related by reflection about the $x$ axis, it follows from Eq.~\eqref{eq:SOI} that $\bm R_2 = \bm R_y^\pi \bm R_1 \bm R_y^\pi$. Alternatively, if they are related by inversion in the $(xy)$ plane, then $\bm R_2 = \bm R_z^\pi \bm R_1^{-1} \bm R_z^\pi$. In both cases, the condition on $\bm R_1$ and $\bm R_2$ obtained above gives $\bm R_1 = \bm R_z^{\pm\pi/2}$ and $\bm R_2 = \bm R_z^{\mp\pi/2}$.

As an example, we consider a network in which two links form a rhombus with an interior angle $2\phi$. Its sides are straight channels of length $a$ and width $w$, as shown in Fig.~\ref{fig:pars}(a). The spin-caging condition then yields
\begin{align}\label{eq:abP1}
&\phi = \arctan \left[(\sqrt{2}\mp 1) \left|\frac{\alpha-\beta}{\alpha+\beta} \right| \right] \,,\\ \label{eq:abP2}
&a = (p+\tfrac12) \frac{\pi\hbar^2}{m^*} \frac{\sqrt{\alpha^2+\beta^2\pm \sqrt{2}\alpha\beta}}{|\alpha^2-\beta^2|} \,,
\end{align}
where $p \geq 0$ is an integer. Geometrically, this corresponds to a spin rotation by $\pi$ (modulo $2\pi$) along each side of the rhombus and an angle of $\pi/4$ between the rotation axes of adjacent sides. Equations~\eqref{eq:abP1}-\eqref{eq:abP2} give the geometric network parameters required to realize spin caging for known spin-orbit parameters. Conversely, for given geometric parameters, the required spin-orbit parameters are $\alpha, \beta  =   (\pi\hbar^2/m^*a)(p+\tfrac12)\sin (\frac{\pi}8 \pm \phi  +\frac{\pi p'}{4})/\sin 2\phi$, where the $+$ ($-$) sign corresponds to $\alpha$ ($\beta$) and $p$, $p'$ are integers.
The relation between the geometric and spin-orbit parameters is illustrated in Fig.~\ref{fig:pars}(b,c).

\section{Spin caging in 1D network}

Now we consider in detail spin diffusion in the 1D chain of rhombi shown in Fig.~\ref{fig:pars}(a). As an example, we choose $\phi = \pi/4$, for which the rhombi become squares, and take the spin-orbit constants corresponding to spin caging, $|\alpha| = \sqrt{2-\sqrt{2}}\pi\hbar^2/(4m^*a)$ and $|\beta| = \sqrt{2+\sqrt{2}}\pi\hbar^2/(4m^*a)$, see red dot in Fig.~\ref{fig:pars}(b,c).

We start by calculating the decay rates of the spin-diffusion eigenmodes. To do so, we consider the unit cell of the chain, a single rhombus, impose Bloch boundary conditions, and calculate the eigenvalues of the right-hand side of Eq.~\eqref{eq:diff}~\cite{Poshakinskiy2025}. The dependence of the eigenmode decay rates on the quasi wave vector along the chain, $q_x$, is shown in Fig.~\ref{fig:disp}. Color encodes the polarization of the modes at the network vertices.
As expected, the dispersion of the mode polarized along $z$ at the vertices (solid red line) is close to quadratic, corresponding to free diffusion. The modes polarized along $x$ and $y$ at the vertices (solid blue and green lines) have almost flat dispersion, indicating their localization. The dashed lines in Fig.~\ref{fig:disp} show the idealized dispersion in the limit of narrow channels, $w \to 0$, and no cubic Dresselhaus spin-orbit interaction, $\beta_3 \to 0$. Applying the standard de Gennes--Alexander approach~\cite{Alexander1983,Rammal1983,Doucot1986,Poshakinskiy2026arxiv} to the spin-diffusive network then yields the (folded) dispersion $\Gamma_{z} =D_s q_x^2 \cos^2 \phi $ for the mode polarized along $z$ at the vertices (red dashed line) and $\Gamma_{x,y} = D_s(\pi/4a)^2$ for the lowest modes polarized along $x$ and $y$ (blue and green dashed lines). There exist also modes with vanishing spin polarization at the vertices (black solid lines) which also have almost flat decay rate dispersion close to $\Gamma_{\rm loop} =  D_s(\pi/2a)^2$ (black dashed line). Such modes are not specific to the present setup: they occur in any network with loops composed of edges of commensurate lengths~\cite{bookBerkolaiko2012}. Due to their stronger decay rate $\Gamma_{\rm loop}=4\Gamma_{x,y}$, these modes are not important for the effects considered in the present paper.

\begin{figure*}[tb!]
    \centering
  \includegraphics[width=0.99\textwidth]{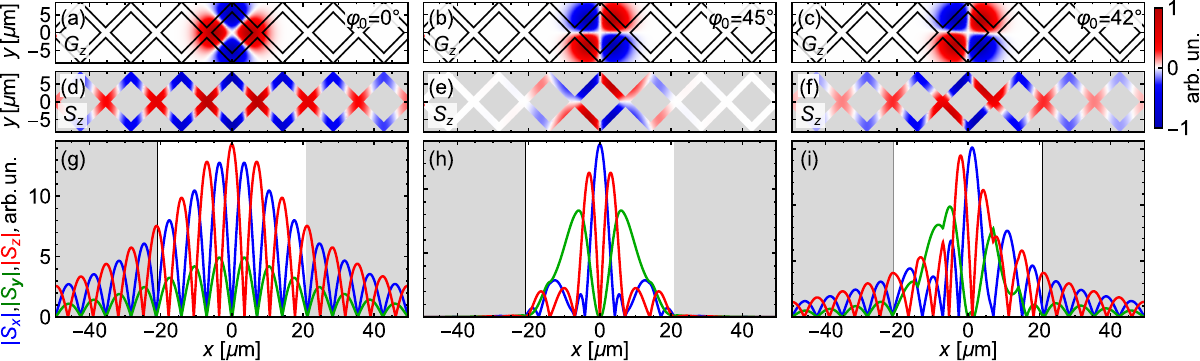}
    \caption{
   Spin diffusion in the 1D rhombic network under continuous spin pumping with a vector vortex beam, Eq.~\eqref{eq:VVB}, with $l=1$ and $r_0=7\,\mu$m. The three columns correspond to the values of $\varphi_0$ indicated in the plot.
(a)--(c) Spatial distribution of the spin generation $G_z$ induced by the beam. (d)--(f) Distribution of the stationary spin polarization $S_z$. (g)--(i) All three components of the stationary spin polarization, evaluated in the middle of the channels in the upper half of the network.
Network parameters are the same as in Fig.~\ref{fig:disp}.}
    \label{fig:VVB}
\end{figure*}

Now we consider spin diffusion in real space to demonstrate the selective caging effect directly. We suppose that spins are pumped locally at a network vertex and investigate how they spread. For simplicity, we consider continuous generation and study the resulting stationary spin distribution; the time-resolved problem leads to similar conclusions. If spin polarization along $z$ is pumped (right column in Fig.~\ref{fig:dif1}), the stationary spin distribution spreads far from the initial excitation spot, shown by the green circle. This corresponds to excitation of the almost free diffusion mode with quadratic dispersion, shown by the red curve in Fig.~\ref{fig:disp}. In contrast, if spin polarization along $x$ or $y$ is pumped at a vertex (left and middle columns in Fig.~\ref{fig:dif1}, respectively), the spread of the spin distribution appears to be limited to one unit cell in either direction. Beyond that, the spin polarization vanishes. It is strictly zero in the limit of the idealized network with $w\to 0$ and a $\delta$-function pump profile.

In practice, using a circularly polarized focused beam, only the $z$ spin component can be generated~\cite{Kohda2017,Passmann2019}. However, spin generation can have a rather arbitrary spatial distribution determined by the pump beam profile. The selective caging described above then leads to the following filtering effect: after diffusing one unit cell beyond the region of spin generation, the spin distribution must assume the spatial pattern corresponding to the almost free mode, shown in the right column of Fig.~\ref{fig:dif1}, because the other modes are caged.

To demonstrate this, we consider optical spin excitation by a vector vortex beam~\cite{Ishihara2023}. In such beams, the degree of circular polarization and the spin generation rate $G_z$ oscillate as functions of the polar coordinate:
\begin{align}\label{eq:VVB}
G_z \propto \e^{-2r^2/r_0^2} r^{2l}\cos [2l(\varphi-\varphi_0)] \,.
\end{align}
Here, the integer $l>0$ is the beam order, $r_0$ characterizes the beam width, $r^2=x^2+y^2$, and $\varphi = \arctan (x/y)$.
Excitation of the spin modes by such beams, with order $l=1$ and the vortex located at the center of a plaquette, is shown in Fig.~\ref{fig:VVB}. Panels (a)--(c) show the spatial distribution of the spin generation $G_z$ for various $\varphi_0$, and the lower panels show the corresponding stationary spin distributions. If $\varphi_0=0$ (left column in Fig.~\ref{fig:VVB}), the pattern of the stationary spin distribution in Fig.~\ref{fig:VVB}(d),(g) matches that in Fig.~\ref{fig:dif1}(c),(f),(i), indicating that the free mode is efficiently excited. If $\varphi_0=\pi/4$ (middle column in Fig.~\ref{fig:VVB}), the spin generation pattern in Fig.~\ref{fig:VVB}(b) is close to the $S_z$ distribution in the $G_y$ caged mode; see Fig.~\ref{fig:dif1}(h). Therefore, the resulting spin distribution also appears to be confined in space: after propagating one plaquette to the left or right of the central plaquette, where the generation is localized, the spin polarization vanishes; see the regions marked in gray in Fig.~\ref{fig:VVB}(h). Finally, if $\varphi_0$ is slightly detuned from $\pi/4$ (right column in Fig.~\ref{fig:VVB}), the filtering effect is realized. Both the free and caged modes are generated, but only the free mode can diffuse beyond one plaquette from the generation area. As a result, the spin distribution in the central area is somewhat arbitrary, but in the regions marked in gray, it matches the pattern of the free mode; in particular, the spin polarization at the vertices is strictly along $z$.

\section{Spin caging in 2D network}

\begin{figure}[tb!]
    \centering
  \includegraphics[width=0.999\columnwidth]{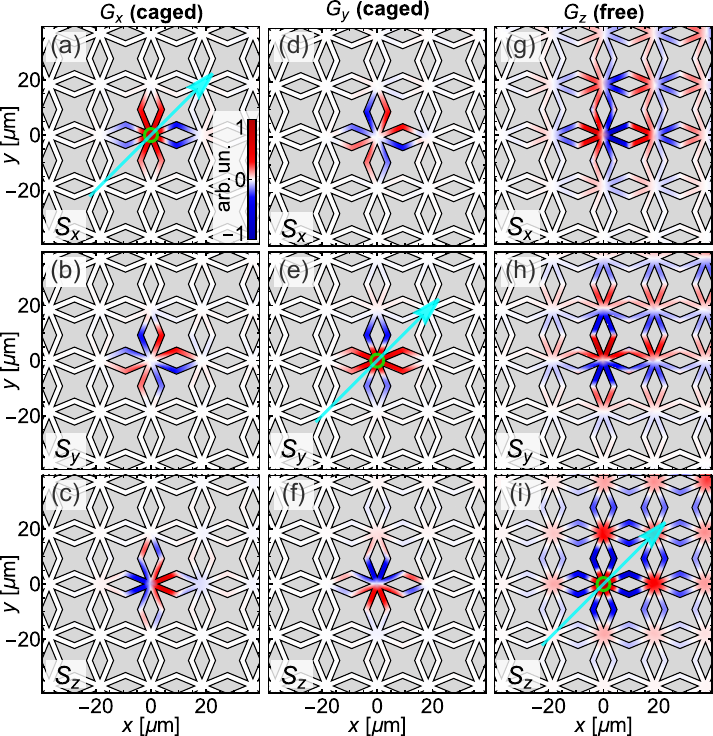}
    \caption{
    Stationary spin distributions in the 2D rhombic network in the presence of an in-plane electric field indicated by the cyan arrow.  Columns correspond to continuous spin pumping of the different spin components $G_x$, $G_y$, and $G_z$ at the center of the plot. Rows show the spatial distributions of the spin-density components $S_x$, $S_y$, and $S_z$ for those cases.
The pump profile was chosen to be Gaussian, with the half-maximum indicated by the green circles.
Network parameters correspond to the red square in Fig.~\ref{fig:pars}(b,c): $\phi=22.5^\circ$, $a=10\,\mu$m, $\alpha = 0$, and $\beta=-1.79\,$meV$\cdot$\AA. The electric field is $\mu E = 2\,\mu$m/ns and is directed $45^\circ$ with respect to the axes.  Other parameters are the same as in Fig.~\ref{fig:disp}.}
    \label{fig:2DE}
\end{figure}

Finally, we consider a generalization of the rhombic network that can realize spin propagation or caging in all in-plane directions. To this end, we consider a 2D square network where every edge is replaced by a rhombus with $\phi=\pi/8$; see Fig.~\ref{fig:2DE}. The spin-orbit parameters that realize selective spin caging in such a network correspond to the red square in Fig.~\ref{fig:pars}(b),(c) and are $\alpha = 0$ and $|\beta| = \pi\hbar^2/(2m^*a)$, or vice versa. For these parameters, the selective spin-caging condition is fulfilled simultaneously for horizontally and vertically oriented rhombi with the same geometric parameters. For generic values of $\alpha$ and $\beta$, the geometric parameters of the horizontally and vertically oriented rhombi need to be tuned separately to fulfill the caging condition.

To demonstrate features of the spin transport in such a network, we additionally apply an in-plane electric field $\bm E$, which leads to the drift of the spin polarization. The drift velocity $\bm v_{\rm d}(\bm r)$ varies across the grid. To determine it, we assume that the electron mobility $\mu$ is constant across the structure and solve Laplace's equation $\nabla^2 \psi = 0$ for the electrostatic potential $\psi(\bm r)$, with Neumann boundary conditions at the channel edges and $\psi = \pm Ed/2$ imposed on two parallel lines separated by a large distance $d \gg a$. We then determine the drift velocity distribution as $\bm v_{\rm d} = \mu \bm\nabla \psi$. To first order in $\bm v_{\rm d}$, spin drift and diffusion are described by Eq.~\eqref{eq:diff} with the following additional terms on the right-hand side~\cite{Kohda2017,Passmann2019}:
\begin{align}\label{eq:drift}
\left.\frac{\partial\bm S}{\partial t}\right|_{\rm drift}
={}&-\sum_{\nu=x,y}v_{{\rm d},\nu}
\left(\frac{\partial}{\partial r_\nu}-\bm\Lambda_\nu\right)\bm S
+\bm\Omega_{\rm d}\times\bm S\,.
\end{align}
Here, the first term describes advection and spin rotation by the first angular harmonic of the spin-orbit field, while $\bm\Omega_{\rm d}=-(2m^*\beta_3/\hbar^2)(v_{{\rm d},x}\bm e_y+v_{{\rm d},y}\bm e_x)$ describes additional spin precession due to the $k$-cubic Dresselhaus spin-orbit interaction~\cite{Altmann2016,Ferreira2017}. At finite $\bm\Omega_d$, the spin rotation angle for a propagating electron depends not only on the trajectory but also on the propagation time, spoiling the caging condition. However, $\beta_3$ is usually much smaller than $\beta$, and the effect of $\bm\Omega_d$ can be neglected. Another important condition for the caging to occur in the presence of the drift is that the spins coming from alternative paths must spend in the vertices enough time to mix, which requires $v_{\rm d} \ll D_s/w$. 

The spin distributions in Fig.~\ref{fig:2DE} show the stationary distributions of all spin components under continuous spin pumping at the center of the plot. When the $x$ or $y$ spin component is pumped at the central vertex, Fig.~\ref{fig:2DE}(a)--(c) and (d)--(f), the spin distribution remains localized within the four rhombi connected to that vertex. The applied electric field leads to no drift, only slight spin redistribution within the four rhombi. In contrast, when the $z$ spin component is pumped, Fig.~\ref{fig:2DE}(g)--(i), the spin distribution drifts freely in the direction of the electric field, indicated by the cyan arrow. Therefore, the proposed structure realizes spin-selective drift. As in the 1D network, the free and localized modes can be efficiently excited by structured light, provided that the distribution of the degree of circular polarization across the beam matches the distribution of the $S_z$ component of the spin-diffusion mode.

\section{Conclusions}

We have demonstrated that, by tuning the geometric and spin-orbit parameters of a semiconductor network, one can satisfy the spin-caging conditions for two spin components. This produces flat decay-rate bands for the caged modes and a quadratic dispersion for the remaining free mode. Both caged and free modes can be excited by localized optical pulses or by structured light with spatially varying circular polarization. After excitation, as spins diffuse or drift through the network under an external electric field, the caged modes remain confined within one unit cell of the excitation spot, whereas the free mode spreads beyond this limit, enabling perfect spin filtering.
The proposed structures can be readily realized by etching a semiconductor quantum well, using fabrication methods already demonstrated for straight quasi-one-dimensional channels~\cite{Altmann2014,Ishihara2014Wires,Grobecker2025}.\\[.5cm]

\section*{Acknowledgments}

We thank M. Kohda and R. Lodo for their critical reading of the manuscript and helpful comments. This work was supported by the RSF
project 23-12-00142-Continuation.

\end{document}